\documentclass[aps,prl,twocolumn]{revtex4}
\usepackage{graphicx}

\begin{document}

\title{Ground-state properties of a one-dimensional system of dipoles}
\author{A.S.~Arkhipov$^{(1)}$, G.E.~Astrakharchik$^{(1,2)}$, A.V.~Belikov$^{(1)}$,
and Yu.E.~Lozovik$^{(1)}$\email{lozovik@isan.troitsk.ru}
}

\affiliation{
{\small\it $^{1}$ Institute of Spectroscopy, 142190 Troitsk, Moscow region, Russia\\
$^{2}$Dipartimento di Fisica, Universit\`a di Trento, and BEC-INFM, I-38050 Povo, Italy }}
\date{\today}

\begin{abstract}
A one-dimensional (1D) Bose system with dipole-dipole repulsion is studied at zero
temperature by means of a Quantum Monte Carlo method. It is shown that in the limit
of small linear density the {\it bosonic} system of dipole moments acquires many
properties of a system of non-interacting {\it fermions}. At larger linear densities
a Variational Monte Carlo calculation suggests a crossover from a liquid-like to a
solid-like state. The system is superfluid on the liquid-like side of the crossover
and is normal in the deep on the solid-like side. Energy and structural functions
are presented for a wide range of densities. Possible realizations of the model are
1D Bose atom systems with permanent dipoles or dipoles induced by static field or
resonance radiation, or indirect excitons in coupled quantum wires, etc. We propose
parameters of a possible experiment and discuss manifestations of the
zero-temperature quantum crossover.
\end{abstract}
\maketitle

Up until now Bose-Einstein condensation has been realized in many different atom and
molecule species with short-range interactions. At low temperatures such an
interaction can be described by a $s$-wave scattering length and is commonly
approximated by a contact pseudopotential. In contrast, some recent work has focused
on the realization of dipole condensates
\cite{experiment1,experiment2,experiment3,experiment4,Lozovik99}. In these systems, the
dipole-dipole interaction extends to much larger distances and significant
differences in the properties (such as the phase diagram and correlation functions)
are expected. Another appealing aspect of a system of dipole moments is the relative
ease of tuning the effective strength of interactions \cite{Goerlitz} which makes
the system highly controllable. Dipole particles are also considered to be a
promising candidate for the implementation of quantum-computing
schemes\cite{Jaksch,Brennen,DeMille}.

On the theoretical side dipole condensates have been mainly studied on a
semiclassical (Gross-Pitaevskii) \cite{Lushnikov} or Bogoliubov \cite{Duncan2}
level. A model Bose-Hubbard Hamiltonian has been used to describe a dipole gas in
optical lattices and a rich phase diagram was found \cite{GoralBaranov,Lozovik99}.
So far there have been no full quantum microscopic computations of the properties of
a homogeneous system.

Recently the Monte Carlo method was used to study helium and molecular hydrogen in
nanotubes\cite{KrotscheckBoronat}. Such a geometry, which is effectively one
dimensional, leads to completely different properties compared to a
three-dimensional sample.

We consider $N$ repulsive dipole moments of mass $M$ located on a line. The
Hamiltonian of such a system is given by
\begin{eqnarray}
\hat H =
-\frac{\hbar^2}{2M}\sum\limits_{i=1}^N\frac{\partial^2}{\partial z_i^2}
+\frac{C_{dd}}{4\pi}\sum\limits_{i<j}\frac{1}{|z_i-z_j|^3}
\label{H}
\end{eqnarray}

We keep in mind two different possible realizations:

1) Cold bosonic atoms, with induced or static dipole momenta, in a transverse trap
so tight that excitations of the levels of the transverse confinement are not
possible and the system is dynamically one-dimensional. The dipoles themselves can
be either induced or permanent. In the case of dipoles induced by an electric field
$E$ the coupling has the form $C_{dd} = E^2\alpha^2$, where $\alpha$ is the static
polarizability. For permanent magnetic dipoles aligned by an external magnetic field
one has $C_{dd} = m^2$, where $m$ is the magnetic dipole moment. The interaction
between the atoms, apart from the dipole forces, contains a short-range scattering
part which is conveniently described by the $s$-wave scattering length $a$. Usually
this contribution is large compared to the dipole force, but recent progress in
applying Feshbach resonance techniques to tune $a$, and even to make it zero, opens
exciting prospects of obtaining a system with purely dipole-dipole interactions. The
strength of the effective coupling $C_{dd}$ can be tuned by changing the electric
field in the case of induced dipoles, and the special technique of fast rotation of
the electric or magnetic field can be applied to permanent dipoles\cite{Goerlitz}.

2) Spatially indirect excitons in two coupled quantum wires. A quantum wire is a
semiconductor nanostructure where an electron or a hole is allowed to move only in
one direction and excitations of the transverse quantization levels are negligible.
In two parallel quantum wires, one containing only holes, and the other only
electrons, holes and electrons couple forming indirect excitons. If such a system is
dilute enough, it constitutes a $1D$ set of dipoles. In this case $C_{dd}
= e^2D^2/\varepsilon$, where $e$ is an electron's charge, $\varepsilon$ is the
dielectric constant of the semiconductor, and $D$ is the distance between the
centers of the quantum wires. This system is 1D counterpart of 2D indirect exciton
system in coupled quantum wells, which was extensively studied both theoretically
\cite{excitonstheory,LozovikYudson,Littlewood,ExcitonPapers},\cite{Lozovik99} and
experimentally\cite{excitonsexper}. The properties of 1D and 2D systems differ
essentially (see below Tonks-Girardeau regime, {\it etc}.), thus an experimental
study of the 1D system is of a great importance.

The Hamiltonian (\ref{H}) can be written in dimensionless form by expressing all
distances $\tilde z = z/r_0$ in units of $r_0 = M C_{dd}/(2\pi\hbar^2)$ and energies
in units of ${\cal E}_0 = \hbar^2/Mr_0^2$. All properties of such a system are
defined by the dimensionless parameter $n r_0$ with $n = N/L$ being the linear
density and $L$ the size of the system.

Our aim is to determine the ground state energy in wide range of densities and to
check the evidence of a possible quantum crossover. In order to define the
structural properties we calculate the pair distribution function (PDF). In terms of
the many-body ground state wave function of the system $\Psi_0$ PDF is written as
\begin{equation}
g_2(|z_1\!-\!z_2|) = \frac{N(N-1)}{n^2}
\frac{\int |\Psi_0(z_1,..,z_N)|^2 dz_3 .. dz_N}
{\int |\Psi_0(z_1,..,z_N)|^2 dz_1 .. dz_N}
\label{g2}
\end{equation}

The static structure factor is directly related to the pair distribution function
\begin{equation}
S(k) = 1 + n\int e^{ikz} [g_2(z)-1]\, dz
\label{Sk}
\end{equation}

The technique of Bragg scattering provides access to the static structure factor in
experiments on bosonic atoms. For the system of indirect excitons measurements of
spatial structure of photoluminescence can give information about possible
crystallization, etc.

We apply Diffusion Monte Carlo (DMC) method, which is one of the most efficient
theoretical tools for investigating zero temperature properties of quantum
systems\cite{Boronat}. We choose the many-body guiding trial wave function $\Psi_T$
in the Bijl-Jastrow form consisting of one- and two- body terms:
\begin{eqnarray}
\Psi_T(z_1, ..., z_N) = \prod\limits_{i=1}^N f_1(z_i) \prod\limits_{j<k} f_2(|z_j-z_k|)
\label{Psi_T}
\end{eqnarray}

To describe the liquid-like side of the crossover it is sufficient to have only the
term describing pair-correlations. We choose it in the form $f_2(|z|) =
\exp\{-[A/(n|z|)]^B\}$, where $A$ and $B$ are variational parameters which we
optimize by minimizing variational energy by carrying out Variational Monte Carlo
(VMC) calculation. As will be discussed below, in the low density limit the wave
function asymptotically approaches the Tonks-Girardeau gas wavefunction
$f_2^{TG}(|z|) = |\sin(\pi z/L)|$ \cite{Girardeau,corr funct}. We checked that in
this regime there is no large difference between using $f_2^{TG}(|z|)$ and
$f_2(|z|)$ even on a variational level. Thus our choice of a two-body term is well
suited to describing a liquid even in the strong mean-field regime (see \cite{corr
funct}).

We account for the spatial quasi-crystalline order by considering Gaussians with
width $C$ for each particle near to a corresponding lattice site $f_1(z_i) =
\exp\{-[n(z_i-z_i^c)]^2/(2C^2)\}$. The sites $z_i^c$ are equally spaced with the
distance $n^{-1}$ and the variational parameter $C$ is defined by VMC optimization.
For the simulation at high density we keep the same type of two-body term $f_2$ as
on the liquid-like side of the crossover, although the values of optimal parameters
may differ.

\begin{figure}
\begin{center}
\includegraphics*[width=0.6\columnwidth,angle=-90]{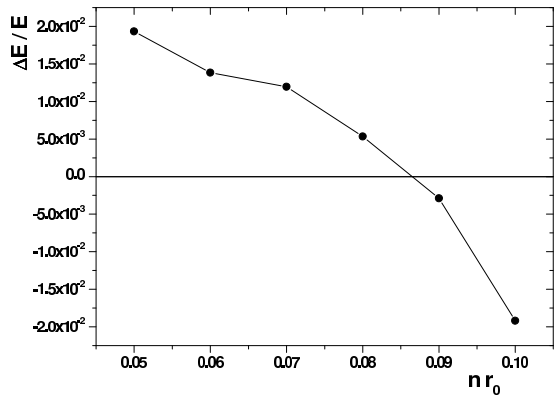}
\caption{Difference between energy calculated using liquid-like and solid-like
wavefunctions $(E_{solid}-E_{liquid})/E$ as a function of the dimensionless
density.}
\label{Fig1}
\end{center}
\end{figure}

In Fig.~\ref{Fig1} we show the comparison of the VMC energy of the liquid-like and
solid-like wavefunctions in the $nr_0 = 0.01-0.1$ density range. In this region the
optimal parameters are $A = 1.6$, $B = 0.4$, $C = 1.16$. We discover that for
densities smaller than $n_c \approx 0.085 r_0^{-1}$ the liquid-like wavefunction
description energetically favorable, although at larger densities the ground state
of the system is better described by a solid-like {\it ansatz}. On the contrary to 3D
and 2D liquid-solid phase transitions, the energetic difference in 1D system is very
small. This suggests that the transition in a one-dimensional system at zero
temperature is of a crossover type.

At a very small density one expects that: (1) in the process of two-body collisions
particles always get reflected back due to the repulsive interaction which is very
intense at short distances, (2) particles stay far apart most of the time, so that
the potential energy of the interaction is small compared to the kinetic energy.
From above it follows that the system is equivalent to a gas of impenetrable bosons
(Tonks-Girardeau gas). It was shown in \cite{Girardeau} that the wave function of
such a Bose system can be mapped onto the wave function of a system of
non-interacting spinless {\it fermions}. The bosonic system acquires many
fermion-like properties (fermionization): the energy is the Fermi energy $E^{TG}/N =
{\cal E}_0\pi^2(nr_0)^2/6$, and the pair distribution function exhibits Friedel-like
oscillations $g^{TG}_2(z) = 1 -\sin^2(\pi n z)/(\pi n z)^2$.

We present the dependence of the energy per particle on the density in
Fig.~\ref{Fig1}. At small density $nr_0\ll 1$ the energy is the same as that of the
Tonks-Girardeau gas $E^{TG}$, which signals fermionization of the system in this
regime. At large density the particles are localized at lattice sites with the
potential energy $E^{str.int.}/N = {\cal E}_0\zeta (3)(nr_0)^3$ being dominant. In
this regime the density dependence of the energy is very strong. Indeed it is cubic,
in contrast to the linear dependence on the mean field for the pseudopotential
interaction and the quadratic dependence for an exactly solvable model with a
$1/z^2$ interaction \cite{Sutherland}. This strong dependence comes from the
diverging short-range behavior of the dipole-dipole interaction. Although in 3D the
dipole-dipole interaction is long-range, this is no longer true in the
one-dimensional case (see, for example, \cite{Ruffo}) and no special techniques like
Ewald summation are required.

We studied the dependence of the energy on the size of the system. The energy has
two contributions: one coming from summation over pairs separated by a distance
smaller than $L/2$ (this contribution is a direct output of the Monte Carlo
calculation) and a tail energy, which is estimated by approximating the density at
$L/2$ by the asymptotic bulk value. We find that the energy per particle as a
function of the system size quickly saturates to its thermodynamic value and the
results obtained for $N=50,100,200$ particles agree within the statistical accuracy
present in our calculation. We also find that the ``tail energy'' contribution is
smaller than $0.2\%$ of the total energy. All reported results are obtained using
$N=100$ particles.
\begin{figure}
\begin{center}
\includegraphics*[angle=-90,width=0.8\columnwidth]{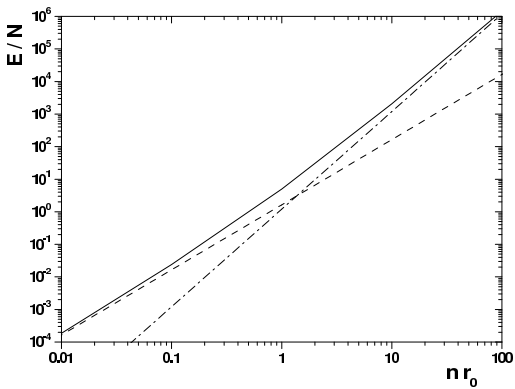}
\caption{Energy per particle as a function of the dimensionless
density (solid line), energy $E^{TG}$ of the Tonks-Girardeau limit
(dashed line), potential energy in the strongly interacting limit
(dash-dotted line). Everything is measured in units of
$\hbar^2/mr_0^2$.}
\label{Fig2}
\end{center}
\end{figure}

\begin{figure}
\begin{center}
\includegraphics*[angle=-90,width=0.8\columnwidth]{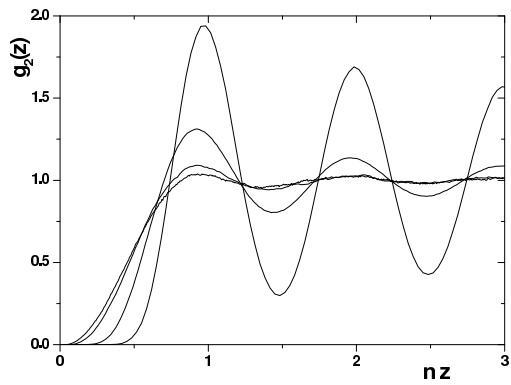}
\caption{Pair distribution function (\ref{g2}) obtained from a DMC
calculation for densities $nr_0 = 10^{-3}, 0.1, 1, 10$ (larger
densities have higher peaks).}
\label{Fig3}
\end{center}
\end{figure}

The pair distribution function (Eq.~(\ref{g2})) is presented in Fig.~\ref{Fig3} for
a range of densities covering all the regions of interest. In a dilute system we
find that amplitude of the oscillation decays rapidly which is characteristic for a
liquid. In particular, at the density $nr_0 = 10^{-3}$ it is almost impossible to
distinguish the pair distribution function from that of spinless fermions, which is
in agreement with the arguments given above. By increasing the density the
oscillation becomes more pronounced. Around the critical density $n_cr_0\approx
0.08$ we enter the solid-like side of the crossover. Further, at larger densities
$nr_0 = 1, 10$ we see manifestations of the localization of particles near lattice
sites.

By performing the Fourier transformation (\ref{Sk}) we obtain the static structure
factor from the pair-distribution function. Knowledge of $S(k)$ is of high
importance as it can be accessed experimentally by using Bragg spectrometry. At
small densities the low-momentum region has a linear form which is related to the
speed of sound $c$ by the Feynman relation $S(k) = \hbar|k|/(2Mc)$. In particular,
in the Tonks-Girardeau limit the static structure factor takes an extremely simple
form: the linear growth matches the asymptotic constant at the wave vector $|k|
= 2\pi n$. Increasing the density leads to the formation of a peak
structure. Presence of the peak in the $S(k)$ is a consequence of the dipole-dipole
interaction, as the peak is absent in a system with only $s$-wave
scattering\cite{Lieb}. In the regime of large density we see several peaks at
integer multiples of $2\pi n$.
\begin{figure}
\begin{center}
\includegraphics*[width=0.8\columnwidth]{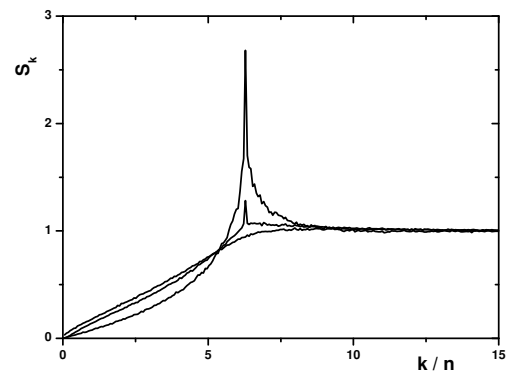}
\caption{Static structure factor obtained from a DMC calculation for
densities $nr_0$ = $10^{-3}$, $0.1$, $1$. A higher first peak
corresponds to a higher density.}
\label{Fig4}
\end{center}
\end{figure}

In order to test superfluidity of the system we use the winding-number
technique\cite{Ceperley}. The superfluid fraction is given by the ratio of the
imaginary time diffusion constants of the center of mass and of a free particle
\cite{disorder}. In contrast to the calculation of the energy, here the symmetry of
the trial wavefunction is crucial. We restore symmetry in the trial wavefunction in
the solid-like region by performing a summation over all sites $f_1(z_i) =
\sum_j\exp\{-[n(z_i-z_j^c)]^2/(2C^2)\}$. We see negligible differences in the energy
(similarly to that found in $^4$He simulations \cite{CeperleyBook}) which justifies
our previous choice of $\Psi_T$. We find that the system is superfluid on the
liquid-like side of the crossover and is normal deep on the solid-like side.

We argue that the critical densities of the quantum crossover can be readily reached
in experiments with interacting indirect excitons in two quantum wires. Let us take
the GaAs parameters ($\varepsilon=12.5$, electronic mass $m_e=0.07m_0$, hole mass
$m_h=0.15m_0$, with $m_0$ being the free electron's mass) for reference. Then the
mass of an exciton is $0.22m_0$. Using quantum wires with a separation between their
axes, $D$, equal to $5\,nm$, we obtain $r_0\approx10^{-6}cm$ and ${\cal E}_0\approx
3meV$. Then the dimensionless density $nr_0=0.08-0.09$ corresponds to an achievable
linear excitonic density $n\approx 10^5cm^{-1}$. The liquid-like and solid-like
regions and the crossover between them should be realized by changing the density of
excitons in the quantum wires.

Interactions between chromium atoms can be efficiently tuned as proposed in
\cite{Goerlitz}. Chromium has the advantage of having a large permanent magnetic
moment when compared to the other alkali atoms. Research into achieving
Bose-condensation in chromium is now a hot
topic\cite{experiment1,experiment2,experiment3,experiment4}. In permanent dipole
chromium atoms the ratio between the strength $C_{dd}$ of the dipole-dipole
interaction and $s$-wave coupling constant is $0.27$ for $^{52}$Cr. Manipulation of
induced electric dipoles is more difficult, although in such a system a ratio of the
order of $10^2$ can be reached leading to the realization of an almost pure dipole
system.

In conclusion we have investigated the ground state properties of a system of dipole
moments by means of a quantum Monte Carlo method. We have found the presence of a
quantum crossover: at small linear densities $nr_0$ system stays liquid-like and
superfluid, although in a more compressed system the solid-like description is
energetically favorable. We have calculated the experimentally accessible pair
distribution function and the static structure factor for a wide range of densities.
In the dilute limit this Bosonic system becomes similar to a system of spinless
fermions (fermionization) and the properties of the system are those of a
Tonks-Girardeau gas. On the liquid-like side of the crossover the system is
superfluid, although it is normal deep on the solid side.

Finally, we have pointed out that the critical density of the quantum crossover can
be achieved in current experiments with interacting excitons in wires and have
proposed the parameters of a possible experimental setup.

We would like to thank F.~Pederiva for useful discussions. G.E.A. acknowledges
support by MIUR. Yu.E.L. is grateful to INTAS and RFBR for support. We thank
B.~Jackson for reading the manuscript.


\begin{references}

\bibitem{experiment1} P.~O.~Schmidt {\it et al.}, Phys. Rev. Lett. {\bf 91}, 193201 (2003)
\bibitem{experiment2} J.~M.~Doyle {\it et al.}, Phys. Rev. A {\bf 65}, 021604 (2002)
\bibitem{experiment3} J.~Stuhler {\it et al.}, Phys. Rev. A {\bf 64}, 031405 (2001)
\bibitem{experiment4} C.~C.~Bradley {\it et al.}, Phys. Rev. A {\bf 61}, 053407 (2000)
\bibitem{Lozovik99} Yu.E. Lozovik, S.Y. Volkov, M. Willander, JETP Lett., {\bf 79},
473 (2004);
Yu.E.Lozovik, S.A.Verzakov, M.Willander, Phys.Lett.A {\bf 260}, 405(1999);
A.I.Belousov, S.A.Verzakov, Yu.E.Lozovik, JETP {\bf 114}, 322 (1998)
\bibitem{Goerlitz} S.~Giovanazzi {\it et al.}
Phys. Rev. Lett. {\bf 89}, 130401 (2002)
\bibitem{Jaksch} D.~Jaksch {\it et al.}, Phys. Rev. Lett. {\bf 85}, 2208 (2000)
\bibitem{Brennen} G.K.~Brennen {\it et al.}, Phys. Rev. A {\bf 65}, 022313 (2002)
\bibitem{DeMille} D.~DeMille, Phys. Rev. Lett {\bf 88}, 067901 (2002)
\bibitem{Lushnikov} P.M.~Lushnikov, Phys. Rev. A {\bf 66}, 051601 (2002)
\bibitem{Duncan2} D.H.J.~O'Dell, S.~Giovanazzi, C.~Eberlein, Phys. Rev. Lett. {\bf 92}, 250401 (2004)
\bibitem{GoralBaranov} K.~G\'oral, L.~Santos, and  M.~Lewenstein, Phys. Rev. Lett. {\bf 88}, 170406 (2002);
M.~Baranov {\it et al.}, Physica Scripta {\bf 102}, 74 (2002)
\bibitem{KrotscheckBoronat} E.~Krotscheck, M.D.~Miller, Phys. Rev. B, {\bf 60}, 13038 (1999);
M.C.~Gordillo, J.~Boronat, and J.~Casulleras, Phys. Rev. B, {\bf 61}, R878
(2000);M.C.~Gordillo, J.~Boronat, and J.~Casulleras, Phys. Rev. Lett., {\bf 85}, 2348 (2000)
\bibitem{excitonstheory} O.L.Berman, Yu.E.Lozovik et.al., Phys.Rev.B{\bf 70}, 5310(2004);
Yu.E.Lozovik et.al., phys. stat. sol. {\bf b241}, 85 (2004);
J.Exp.Theor.Phys. {\bf 98}, 582(2004);
Yu.E.Lozovik {\it et al.}, PRB {\bf 59},5627 (1999);
JETP Lett. {\bf 64}, 573 (1996);
Yu.~E. Lozovik {\it et al.}, JETP Lett. {\bf 69}, 616 (1999);
Yu.~E. Lozovik, A.~V. Poushnov, Phys.~Lett.~{\bf A 228}, 399 (1997);
S.~I. Shevchenko, Phys.~Rev.~Lett. {\bf 72}, 3242 (1994);
Yu.~E.Lozovik {\it et al.}, J. Phys.C, {\bf 14}, 12457 (2002).
\bibitem{LozovikYudson} Yu.E. Lozovik, V.I. Yudson, JETP Lett. {\bf 22}, 26(1975);
JETP {\bf 44}, 389 (1976); Sol. St. Comms. {\bf 21}, 211(1977);
Physica {\bf A93}, 493 (1978)
\bibitem{Littlewood} Xu. Zhu, P.B. Littlewood, M.S. Hybertsen, T.M. Rice,
Phys. Rev. Lett. {\bf 74}, 1633 (1995);
S. Conti, G. Vignale, A.H. MacDonald, Phys. Rev. {\bf B 57}, R6846 (1998);
M.A. Olivares-Robles, S.E. Ulloa, Phys. Rev. {\bf B 64}, 115302 (2001)
\bibitem{ExcitonPapers}
S.~De Palo, F.~Rapisauda, and G.~Senatore, Phys. Rev. Lett. {\bf 88}, 206401 (2002)
\bibitem{excitonsexper}
A.A.Dremin, V.B.Timofeev, A.V.Larionov {\it et al.} JETP lett., {\bf 76}, 450(2002).;
R.Rapaport, G.Chen, D.Snoke {\it et al.}, Phys. Rev.Lett. {\bf 92}, 117405 (2004);
L.V.Butov, L.S.Levitov {\it et al.}, Phys. Rev. Lett. {\bf 92}, 117404 (2004);
V.V.Krivolapchuk, E.S.Moskalenko {\it et al.}, Phys. Rev. {\bf B64}, 045313 (2001)
\bibitem{Boronat} For details on the DMC method see, for example,
J.~Boronat, J.~Casulleras, Phys. Rev. B {\bf 49}, 8920 (1994)
\bibitem{Girardeau} M.~Girardeau, J. Math. Phys. (N.Y.) {\bf 1}, 516 (1960)
\bibitem{corr funct} G. E. Astrakharchik and S. Giorgini, Phys. Rev. A {\bf 66}, 053614 (2002)
\bibitem{Lieb} E.H.~Lieb and W.~Liniger, Phys. Rev. {\bf 130}, 1605 (1963)
\bibitem{Sutherland} B.~Sutherland, J. Math. Phys. {\bf 12} 245 (1971);
F.~Calogero, J. Math. Phys. {\bf 10}, 2191, 2197 (1969)
\bibitem{Ruffo} {\it Dynamics and thermodynamics of systems with long-range interactions},
ed.
T.~Dauxois {\it et al.}, Springer 
2002
\bibitem{Ceperley} E.L. Pollock and D.M. Ceperley, Phys. Rev. B {\bf 36}, 8343 (1987);
S. Zhang {\it et al.}, Phys. Rev. Lett. {\bf 74}, 1500 (1995)
\bibitem{disorder} G.E.~Astrakharchik {\it et al.} Phys. Rev. A {\bf 66}, 023603 (2002)
\bibitem{CeperleyBook} D.M.~Ceperley and M.H.~Kalos, {\it Monte Carlo Methods in Statistical Physics}, edited by K.~Binder, Springer, 1979
\end{references}
\end{document}